\documentstyle{article} 

\begin{document} 
\centerline{\Large A calculable and quasi-practical gas} \vskip 10pt 
\centerline{C. Y. Chen} \centerline{Department of Physics, Beijing University 
of Aeronautics} \centerline{ and Astronautics, Beijing, 100083, P.R.China} 
\centerline{Email: cychen@buaa.edu.cn} \vskip 10pt \begin{abstract} A new 
kinetic approach is developed and a quasi-practical gas is defined to which 
the new approach can be applied. One of the advantages of this new approach 
over the standard one is direct calculability in terms of today's 
computational means. 

\end{abstract} \vskip 5pt \centerline{PACS 51.10.+y - Kinetic and transport 
theory of gases} \vskip 5pt 

\section{Introduction} The impressive progress of nonlinear science in the 
last several decades revealed a large number of structures and mechanisms 
that were novel and fascinating to almost all of us. This has, in a sense, 
suggested that several aspects of our existing statistical theory, whose 
first kinetic equation was proposed more than a century ago, need certain 
kinds of `upgrade' or `renewal'. 

In some of our relatively recent papers{\cite{chen1,chen2}, we were tempted 
to analyze the standard kinetic theory from different perspectives. It was 
argued that (i) in various artificial and realistic situations, distribution 
functions of gases may have very unconventional structures: they can be, for 
instance, discontinuous at every and each spatial point and thus the usual 
differential apparatus becomes inapplicable (after all, there are indeed many 
structures and mechanisms in nature to which the use of the usual 
differential apparatus is quite limited); and that (ii) while the left side 
of the Boltzmann equation is symmetric in terms of partial derivatives with 
respect to the position and velocity, the right side of the equation is 
constructed entirely in the velocity space (the position coordinates merely 
serve as inactive parameters), which is incomprehensible especially in the 
mathematical sense. 

The major objective of this work is, however, rather practical: to define a 
kinetic gas and, at the same time, to find out an algorithm which can be used 
to calculate the behavior of the gas rather completely. It is hoped that if 
such study gets established somehow, further studies, either concerning the 
foundation of the existing statistical theory or concerning the development 
of more general treatment, will be inspired and promoted. 

The structure of this paper is the following. In Sect.~2, we propose our 
working model, a gas leaking out of a container, resembling the cavity model 
for the black-body radiation. In Sect.~3, the zeroth-order distribution 
function of the leaking gas is formulated and, in passing, it is shown that 
distribution functions of real gases may have radically discontinuous 
structures untreatable for the standard theory. In Sect.~4, the collisional 
correction is investigated with help of a methodology that is slightly 
different in form but much different in concept from the standard treatment. 
In Sect.~5, we further comment on why the distribution function averaged over 
finite velocity solid-angle ranges needs to be introduced. Sect.~6 summarizes 
the paper. 

\section{A leaking gas} 

At the kinetic level, one of the customary conceptions about calculating a 
practical gas is to solve the Boltzmann differential-integral equation with 
help of a difference scheme. Known difficulties associated with this 
conception may be summarized as follows. \begin{enumerate} \item There are 
seven variables: time, three spatial coordinates and three velocity 
components. If we divide each variable into $N$ intervals, `the degrees of 
freedom' of the system are $N^7$. To reveal the true properties of a 
nonequilibrium phenomenon, such as those related to a turbulence, $N$ has to 
be rather large and $N^7$ has to be terribly huge, so that no today's 
computational means really helps. \item The complex nature of the collisional 
operator in Boltzmann's equation makes the situation worse. \item Boundary 
conditions and initial conditions impose extra problems. As to a differential 
equation, well specified boundary and initial conditions usually mean that a 
uniquely and clearly defined solution can be constructed, say, from a 
difference scheme. (The uniqueness may not be truly desired in view of the 
fact that bifurcations take place in nature.) As to an integral equation, 
boundary and initial conditions may be specified rather loosely and it leaves 
us certain room to `manipulate'. Boltzmann's equation is a 
differential-integral equation and it is not entirely clear in which 
direction we should go. \end{enumerate} 
Due to these difficulties (and 
possibly many more), almost no realistic problems have been fully treated, 
let alone conclusive comparison between the kinetic theory and realistic 
experiments. 

Interestingly enough, the aforementioned difficulties, at least some of them, 
are not intrinsic to the dynamics that we wish to study. In some sense, if 
the standard approach had not dominated our mind too strongly, we might have 
already had workable alternative approaches, at least for some special cases. 
In what follows, we shall try to substantiate this viewpoint. 

We consider a gas consisting of hard spheres and confined to a closed 
container. The interactions between the walls and particles, and between the 
particles themselves, make the gas solidly and quickly in equilibrium. 
Namely, the probability of finding particles in a phase volume element $d{\bf 
r}d{\bf v}=dxdy dz dv_x dv_y dv_z$ takes its Maxwell form 
\begin{equation} \label{fm0} f_M\equiv n_0\left(\frac m{2\pi\kappa 
T}\right)^{\frac 32} \exp\left( - \frac{mv^2}{2\kappa T}\right), 
\end{equation}
where $m$ is the mass of particle, $\kappa$ the Boltzmann constant, $T$ the 
temperature of the gas and $n_0$ the particle density in the container. 
\vskip15pt 

\hspace{40pt} \setlength{\unitlength}{0.015in} \begin{picture}(200,100) 
\put(30,100){\line(1,0){60}}
\put(30,10){\line(1,0){60}}
\put(40,20){\line(1,0){40}}
\put(40,90){\line(1,0){40}}

\put(90,53){\line(0,-1){43}}
\put(90,57){\line(0,1){43}}
\put(90,57){\line(-2,3){10}}
\put(90,53){\line(-2,-3){10}}

\multiput(30,10)(2,0){6}{\line(0,1){90}}
\multiput(40,10)(2,0){21}{\line(0,1){10}}
\multiput(40,100)(2,0){21}{\line(0,-1){10}}

\put(80,10){\line(0,1){28}}
\put(82,10){\line(0,1){31}}
\put(84,10){\line(0,1){34}}
\put(86,10){\line(0,1){37}}
\put(88,10){\line(0,1){40}}

\put(80,100){\line(0,-1){28}}
\put(82,100){\line(0,-1){31}}
\put(84,100){\line(0,-1){34}}
\put(86,100){\line(0,-1){37}}
\put(88,100){\line(0,-1){40}}

\multiput(42,23)(0,4){17}{\circle*{1}}
\multiput(46,23)(0,4){17}{\circle*{1}}
\multiput(50,23)(0,4){17}{\circle*{1}}
\multiput(54,23)(0,4){17}{\circle*{1}}
\multiput(58,23)(0,4){17}{\circle*{1}}
\multiput(62,23)(0,4){17}{\circle*{1}}
\multiput(66,23)(0,4){17}{\circle*{1}}
\multiput(70,23)(0,4){17}{\circle*{1}}
\multiput(74,23)(0,4){17}{\circle*{1}}
\multiput(78,23)(0,4){17}{\circle*{1}}
\multiput(82,47)(0,4){5}{\circle*{1}}
\multiput(86,51)(0,4){3}{\circle*{1}}

\multiput(95,55)(1.8,0){20}{\circle*{0.7}}
\multiput(94,57)(1.6,0.9){20}{\circle*{0.7}}
\multiput(92.5,58.5)(0.9,1.6){20}{\circle*{0.7}}
\multiput(94,53)(1.6,-0.9){20}{\circle*{0.7}}
\multiput(92.5,51.5)(0.9,-1.6){20}{\circle*{0.7}}\end{picture} 

\vspace{-8pt} \begin{center} \begin{minipage}{10cm} { \vskip-0.3cm Figure~1: 
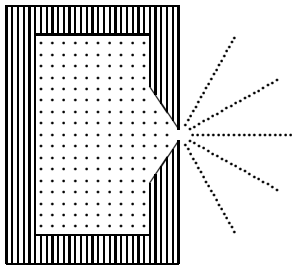
Schematic of a leaking gas.} \end{minipage} \end{center} 

We then suppose that there is a small hole on the wall of the container, as 
shown in Fig.~1. The aim of our formalism is to determine the distribution 
function of the leaking gas. For convenience of discussion, it is further 
assumed that (i) the hole is indeed small so that the leaking is relatively 
slow and the relevant distribution function can be considered to be 
time-independent; (ii) since the regions far from the hole are kept in the 
vacuum state (by a pump for instance), no incoming particles need to be taken 
into account; and (iii) the leaking gas outside the container is so dilute 
that the zeroth-order solution can be determined by the collisionless 
trajectories of particles while the first-order correction can be formulated 
by assuming the particles of the leaking gas to collide with each other once 
and only once. 

\section{The zeroth-order distribution function} 

If we ignore the particle-to-particle interaction and ignore the 
gravitational force acting on each particle, the collisionless motion of 
every particle of the leaking gas is conceptually simple---moving along a 
straight line. However, it is still necessary and interesting to express the 
corresponding distribution function in a proper mathematical form. 

We start our discussion with the conventional theory. The theory states that 
the distribution function satisfies, with collisions ignored completely, 
\begin{equation} \label{bltz} \frac{\partial f}{\partial t}+{\bf v}\cdot 
\frac {\partial f} {\partial {\bf r}}+ {{\bf F}\over m} \cdot \frac {\partial 
f} {\partial {\bf v}}=0,
\end{equation}
where $\bf F$ represents the external force. Equation (\ref{bltz}) is 
sometimes termed the collisionless Boltzmann equation\cite{reif}. A natural 
notion related to such terminology is that any procedure, a difference scheme 
for instance, if applicable in solving the regular Boltzmann equation, must 
be applicable in solving this reduced equation. Strangely, complicated and 
subtle issues arise from this `natural' notion; and we shall discuss these 
issues at the end of this section. 

To formulate the zeroth-order distribution function, the following `slightly 
different' approach is helpful. Rewriting Equation (\ref{bltz}) along a 
particle path in the phase space, we arrive at the path 
invariance\cite{reif,harris} 
\begin{equation} \label{fm}\left. \frac{df}{dt}\right|_{\rm path}=0. 
\end{equation}
As far as our leaking gas is concerned, this means $f|_{\rm path}=f_M$, where 
$f_M$ has been defined by expression (\ref{fm0}). \vskip 20pt 

\hspace{90pt} \setlength{\unitlength}{0.014in} \begin{picture}(200,100) 

\put(109,97){\line(1,-2){5}}
\put(107,79){\makebox(20,8)[l]{$\Delta \Omega$}}
\put(92,89){\makebox(20,8)[l]{$\bf r$}}

\put(42,100){\line(0,-1){25.5}}
\put(39,100){\line(0,-1){21}}
\put(36,100){\line(0,-1){16.5}}
\put(33,100){\line(0,-1){12}}
\put(30,100){\line(0,-1){7.5}}
\put(42,10){\line(0,1){25.5}}
\put(39,10){\line(0,1){21}}
\put(36,10){\line(0,1){16.5}}
\put(33,10){\line(0,1){12}}
\put(30,10){\line(0,1){7.5}}

\put(45,100){\line(0,-1){30}}
\put(45,10){\line(0,1){30}}
\multiput(10,10)(0,5.0){19}{\circle*{1.5}}
\multiput(15,10)(0,5.0){19}{\circle*{1.5}}
\multiput(20,10)(0,5.0){19}{\circle*{1.5}}
\multiput(25,15)(0,5.0){17}{\circle*{1.5}}
\multiput(30,25)(0,5.0){13}{\circle*{1.5}}
\multiput(35,30)(0,5.0){11}{\circle*{1.5}}
\multiput(40,40)(0,5.0){7}{\circle*{1.5}}

\multiput(45,43)(0,6){5}{\line(0,1){3}}
\put(45,50){\line(2,-3){7}}
\put(50,32){\makebox(20,8)[l]{$\Delta S_0$}}

\put(45,70){\line(-2,3){20}}
\put(45,40){\line(-2,-3){20}}
\put(45,55){\vector(3,2){54}}
\multiput(45,55)(3,0){15}{\circle*{1}}
\put(56,55.5){\makebox(20,8)[l]{$\theta$}}

\multiput(47,69)(2,0.85){35}{\circle*{1}}
\multiput(47,42)(1.5,1.4){45}{\circle*{1}}
\end{picture} 

\vspace{-5pt}\begin{center} \begin{minipage}{10cm} { \vskip-0.3cm Figure~2: 
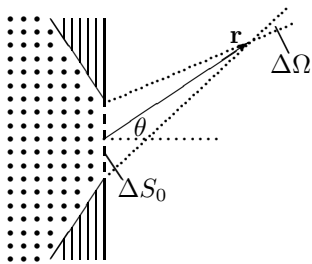
The velocity distribution at the position $\bf r$.} \end{minipage} 
\end{center} 

Then, it seems that the zeroth-order distribution function on the right side 
of the hole is uniformly identical since any spatial point there is reachable 
along a path radiating from the inside of the container (the external force 
is zero and the paths in the phase space and in the spatial space are the 
same). This notion is, however, rather misleading. By moving together with a 
particle of the leaking gas, an observer easily realizes that the true 
particle density around him becomes lower and lower. With this realization in 
mind, it can easily be found that the distribution function of the leaking 
gas is strongly limited in the velocity space: the farther from the hole the 
stronger the limitation is. Namely, as shown in Fig.~2, the true implication 
of (\ref{fm}) is 
\begin{equation} \label{cone} f({\bf r},{\bf v}) =\left\{ \begin{array}{ll} 
f_M &{\rm the{\hskip 4pt} direction{\hskip 4pt} of {\hskip 4pt} {\bf 
v}{\hskip 4pt} within{\hskip 4pt} \Delta\Omega} \\ 0 & {\rm otherwise} 
\end{array} \right. , 
\end{equation}
where $\Delta\Omega$ is a solid-angle range whose size is defined by the 
solid-angle range 
\begin{equation} \label{cone1} \frac{\Delta S_0 \cos \theta}{r^2} \quad{\rm 
with}\quad r=|{\bf r}| 
\end{equation}
and whose direction is defined by the position vector ${\bf r}$ (with the 
origin at the hole). Heuristically and intuitively, it can be said that the 
velocity distribution at every point there is a sting-like function. In 
almost all the regions on the right side of the hole, where $r$ is relatively 
large or $\cos\theta$ is relatively small, the stings tend to be infinitely 
acute. 

Though (\ref{cone}) and (\ref{cone1}) correctly describe the distribution 
function, there are some kinds of inconvenience in applying them in 
analytical calculations. For practical and theoretical reasons, we wish to 
express the distribution function $f({\bf r},{\bf v})$ directly and 
explicitly in terms of ${\bf r}$ and ${\bf v}$. 

To get such expression, we adopt the picture that the sting-like velocity 
distributions of the leaking gas are indeed infinitely acute, either by 
assuming the hole to be truly small or by assuming the interested region to 
be truly distant. Under this understanding, we can use a $\delta$-function to 
reflect the limitation of the distribution function in the velocity space and 
replace (\ref{cone}) and (\ref{cone1}) by 
\begin{equation} \label{delta} f({\bf r},{\bf v})\equiv 
g(r,v)\delta\left(\Omega_{\bf v}- \Omega_{\bf r} \right) 
\end{equation}
with $$ g(r,v) = n_0 \left(\frac m{2\pi\kappa T}\right)^{3/2} e^{- 
\frac{mv^2}{2\kappa T}} \frac{\Delta S_0 \cos\theta}{ r^2}, $$ where 
$\Omega_{\bf v}$ is the solid angle in the direction of $\bf v$ and 
$\Omega_{\bf r}$ is the solid angle in the direction of $\bf r$. Expression 
(\ref{delta}) enables us to do analytical calculation easily. For instance, 
with help of it we may compute the zeroth-order particle density outside the 
container by 
\begin{equation} n({\bf r})=\int f({\bf r},{\bf v}) d{\bf v} =\int f v^2dv 
d\Omega= n_0\frac{\Delta S_0 \cos\theta}{4\pi r^2}, 
\end{equation}
which is quantitatively consistent with our intuitive notion about the 
particle density outside the container. 

Before leaving this subject, we go back to discuss several usual concepts 
concerning the collisionless and regular Boltzmann equations. It is commonly 
believed that the collisionless Boltzmann equation (\ref{bltz}) is completely 
equivalent to the path-invariance expressed by (\ref{fm}). In our view, this 
equivalence is just a formal one. For one thing, expression (\ref{fm}) makes 
sense strictly along a particle path, which is governed by the collisionless 
motion equations of single particle; while Boltzmann's equations are supposed 
to be solved by a difference scheme with help of boundary and initial 
conditions in which trajectories of individual particle do not play any role. 
For another, unconventional topology and dynamics are related to (\ref{fm}). 
Distribution functions described by it, such as (\ref{cone}) and 
(\ref{delta}), can be shaped like stings; more than that, these stings 
`spread' out along paths of particles, become sharper and sharper when 
spreading, and continue to spread out after being infinitely sharp. 
(Distributions produced by boundaries can be even `sharper', see Sect.~5.) 
Boltzmann's equations, being ones containing differential terms, are not 
compatible with these `radical' things. 

\section{The collisional correction} In this section, we try to deal with 
collisions. To make our discussion involve less details, we shall only 
formulate the first-order correction. Namely, it is assumed that the 
particles expressed by (\ref{delta}) will collide once and only once 
(although further extension can be made with no principal difficulty). 

\hspace{-25pt} \setlength{\unitlength}{0.020in} \begin{picture}(200,100) 
 
\put(90,53){\line(0,-1){23}}
\put(90,57){\line(0,1){23}}
\put(90,57){\line(-2,3){10}}
\put(90,53){\line(-2,-3){10}}

\put(80,30){\line(0,1){8}}
\put(82,30){\line(0,1){11}}
\put(84,30){\line(0,1){14}}
\put(86,30){\line(0,1){17}}
\put(88,30){\line(0,1){20}}

\put(80,80){\line(0,-1){8}}
\put(82,80){\line(0,-1){11}}
\put(84,80){\line(0,-1){14}}
\put(86,80){\line(0,-1){17}}
\put(88,80){\line(0,-1){20}}

\multiput(82,47)(0,4){5}{\circle*{1.5}}
\multiput(86,51)(0,4){3}{\circle*{1.5}}

\multiput(95,55)(1.8,0){14}{\circle*{0.7}}
\multiput(94,57)(1.6,0.9){14}{\circle*{0.7}}
\multiput(92.5,58.5)(0.9,1.6){14}{\circle*{0.7}}
\multiput(94,53)(1.6,-0.9){14}{\circle*{0.7}}
\multiput(92.5,51.5)(0.9,-1.6){14}{\circle*{0.7}}

\multiput(150.2,39.4)(16.6,8.2){2}{\line(1,-2){8}}
\multiput(150.2,39.4)(9.2,4.6){2}{\line(2,1){7}}
\put(158,44.25){\line(-1,5){10}}
\put(158,44.25){\line(-4,5){33}}
\put(157,49){\line(-1,0){3}}
\put(156,54){\line(-1,0){5.5}}
\put(155,59){\line(-1,0){8.5}}
\put(154,64){\line(-1,0){11}}
\put(153,69){\line(-1,0){14.3}}
\put(152,74){\line(-1,0){17.5}}
\put(151,79){\line(-1,0){20.5}}
\put(150,84){\line(-1,0){24}}
\put(149,89){\line(-1,0){12}}

\put(126,24){\makebox(20,8)[l]{Detector}}
\put(153,79){\makebox(20,8)[l]{Effective}}
\put(153,72){\makebox(20,8)[l]{cone $-\Delta\Omega$}}
\end{picture} 

\vspace{-29pt}\begin{center} \begin{minipage}{10cm} { \vskip-0.3cm Figure~3: 
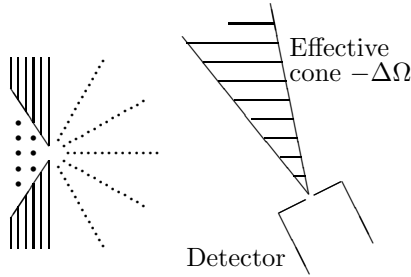
A particle detector is placed in the leaking gas.} \end{minipage} 
\end{center} 

What can be measured in an experiment should be of our first concern. For 
this reason, we consider a particle detector placed somewhere in the leaking 
gas and try to determine how many particles will be recorded by the detector, 
as illustrated in Fig.~3. There are several essential things worth mentioning 
about the detector. \begin{enumerate} \item The opening allowing particles to 
enter the detector is considered to be sufficiently small. Or, in the 
mathematical language, we regard the area of the opening, denoted by $\Delta 
S$, as an infinitesimal quantity throughout this section. \item Without 
specifying concrete structure and mechanism of the detector, it is assumed 
that every particle that enters the detector and is in the velocity range 
\begin{equation} \label{vvolume} \Delta {\bf v}= \Delta v v^2 \Delta \Omega, 
\end{equation}
where $\Delta v$ and $\Delta \Omega$ are predetermined before the 
measurement, will be recorded, and every other particle will not. \item While 
$\Delta S$ and $\Delta v$ are allowed to shrink to zero, we define $\Delta 
\Omega$ as a finite solid-angle range. As will be seen, the discrimination 
against $\Delta\Omega$ is taken almost entirely by necessity. \end{enumerate} 

If the detector works as described above, do we know the distribution 
function at the entry of the detector? The answer to it is almost a positive 
one. If $\Delta N$ is the number that the detector counts during $\Delta t$, 
the particle density in the phase volume element $(\Delta S v \Delta 
t)(\Delta v v^2 \Delta \Omega)$ is 
\begin{equation} \label{deltaN} f(t,{\bf r}, {\bf v},\Delta \Omega )\approx 
\frac {\Delta N}{(\Delta S v \Delta t)(\Delta v v^2 \Delta \Omega)}.
\end{equation}
The form of (\ref{deltaN}) illustrates one of the most distinctive features 
of this approach: it tries to calculate the distribution function directly 
rather than to formulate a dynamic equation concerning how many particles 
enter and leave a phase volume element. Apart from other advantages, this 
brings to us a lot of convenience in computational terms. 

Noticing that only the collisions taking place within the shaded spatial cone 
$-\Delta\Omega$, which is equal and opposite to the velocity cone $\Delta 
\Omega$, can possibly contribute to $\Delta N$, as shown in Fig.~3, we call 
the region inside $-\Delta\Omega$ the effective cone. Since this effective 
cone is a finite one ($\Delta \Omega$ is finite), our primary task is to 
divide it into many subregions, denoted by $d{\bf r}'$ (the origin of the 
coordinates is still at the center of the hole as in Sect.~3), and to 
calculate how collisions in each of the subregions give contribution to 
$\Delta N$. 

Consider that two beams of identical particles (but still distinguishable in 
terms of classical mechanics) 
\begin{equation} f(t',{\bf v}',{\bf r}')d{\bf v}' \quad {\rm and} \quad 
f(t',{\bf v}'_1,{\bf r}')d{\bf v}'_1 
\end{equation}
collide within $d{\bf r}'$ and at time $t'$, producing particles with 
velocities ${\bf v}$ and ${\bf v}_1$ respectively. It is noted that there is 
a time delay 
\begin{equation} t-t'=\frac{|{\bf r}-{\bf r}'|} {v}\quad {\rm with}\quad 
v=|{\bf v}| 
\end{equation}
between the time of collision and the time of detection. Since the leaking is 
assumed to be relatively slow (or the gas inside the container is supplied by 
an external gas source), we think of our problem as a time-independent one 
and pay no more attention to the time variable hereafter. To better observe 
the collisions, we define 
\begin{equation} \label{cu} \left\{\begin{array}{l} {\bf v}'+{\bf 
v}'_1=2{\bf c}'\cr {\bf v}'-{\bf v}'_1=2{\bf u}' \end{array}\right. \quad 
{\rm and}\quad \left\{\begin{array}{l} {\bf v}+{\bf v}_1=2{\bf c}\cr {\bf 
v}-{\bf v}_1=2{\bf u} \end{array}\right. . 
\end{equation}
That is to say, by virtue of the conservation laws of energy and momentum, 
${\bf c}={\bf c}'$ represents the velocity of the center-of-mass and $u=|{\bf 
u}|=|{\bf u}'|$ stands for the particle speed relative to the center-of-mass. 
Examining the beam-to-beam collision in the center-of-mass frame, we find 
that the differential number of the colliding particles is 
\begin{equation} \label{twobeams} [d{\bf r}'f({\bf r}',{\bf v}')d{\bf v}'] 
[f({\bf r}',{\bf v}'_1)d{\bf v}'_1] [2u\sigma_c({\bf u}',{\bf u}) d \Omega_c 
\Delta t ],
\end{equation}
where $\Omega_c$ is the solid angle between ${\bf u}'$ and ${\bf u}$ and 
$\sigma_c$ is the cross section associated with particles emerging within the 
solid-angle range $d\Omega_c$. By integrating (\ref{twobeams}) and getting 
help from (\ref{delta}), the right side of (\ref{deltaN}) is equal to 
\begin{equation} \label{entry} \int_{-\Delta \Omega}d{\bf r}'\int_{\Delta 
v\Delta \Omega_0} d\Omega_c \int_0^\infty dv'\int_0^\infty dv'_1 
\frac{2u\sigma_c g(r',v') g(r',v'_1)} {(|{\bf r}-{\bf r}'|^2 \Delta \Omega_0 
v)\dot (v^2\Delta v \Delta \Omega )}, 
\end{equation}
where $\Delta \Omega_0$ is the solid-angle range formed by the point $d{\bf 
r}'$ (as the apex) and the detector opening $\Delta S$ (as the base), and the 
subindex $\Delta v\Delta \Omega_0$ there states that only particles that can 
be recorded by the detector will be taken into account. Since the size of 
$\Delta \Omega_0$ is `much smaller' than that of $\Delta\Omega$, every 
particle starting its journey from the effective cone and entering the 
detector will be treated as one within $\Delta \Omega$. With help of the 
variable transformation from $(v',v'_1)$ to $(c',u')$ and finally to $(c,u)$, 
we rewrite expression (\ref{entry}) as 
\begin{equation} \label{entry1} \int_{-\Delta \Omega}d{\bf r}'\int_{\Delta 
v\Delta \Omega_0} d\Omega_c \int_0^\infty dc \int_{-c}^{c} du 
\frac{2u\sigma_c\cdot \|J\| \cdot g(c+u) g(c-u)} {(|{\bf r}-{\bf r}'|^2 
\Delta \Omega_0 v)\dot (v^2\Delta v \Delta \Omega )}, 
\end{equation}
in which the Jacobian between the variable transformation is 
\begin{equation} \| J\|=\frac{\partial (v',v'_1)}{\partial (c,u)}=2.
\end{equation}
In view of the symmetry of the cross section there, we have 
\begin{equation} \int_{-c}^c du\cdots =2\int_0^c du \cdots. 
\end{equation}
\vspace{-0.5cm}

\hspace{30pt} \setlength{\unitlength}{0.013in} \begin{picture}(200,135) 

\put(135,93){\makebox(35,8)[l]{${\bf c}$}}
\put(190,50){\makebox(35,8)[l]{$\bf u$}}
\put(220,90){\vector(-1,-1){60}}
\put(70,90){\line(1,0){150}}
\put(217.5,90.5){\vector(1,0){1}}
\multiput(70,90)(3.0,-2.3){32}{\circle*{1.2}}
\multiput(70,90)(3.0,-1.7){35}{\circle*{1.2}}
\multiput(163,18.7)(1.6,2.4){6}{\circle*{1.2}}
\multiput(154,42.4)(-1.6,-2.4){5}{\circle*{1.2}}
\put(70,60){\makebox(35,8)[l]{$\Delta\Omega_0$}}
\put(140,14){\makebox(35,8)[l]{$\Delta v$}}

\put(143.22,53.87){\circle*{1}}
\put(144.80,50.69){\circle*{1}}
\put(146.52,47.57){\circle*{1}}
\put(148.36,44.53){\circle*{1}}
\put(150.32,41.57){\circle*{1}}
\put(152.41,38.70){\circle*{1}}
\put(154.62,35.91){\circle*{1}}
\put(156.94,33.22){\circle*{1}}
\put(159.38,30.63){\circle*{1}}
\put(161.91,28.15){\circle*{1}}
\put(164.56,25.77){\circle*{1}}
\put(167.29,23.50){\circle*{1}}
\put(170.13,21.35){\circle*{1}}
\put(173.04,19.32){\circle*{1}}
\put(176.04,17.42){\circle*{1}}
\put(179.12,15.64){\circle*{1}}
\put(182.27,14.00){\circle*{1}}
\put(185.49,12.48){\circle*{1}}
\end{picture} 

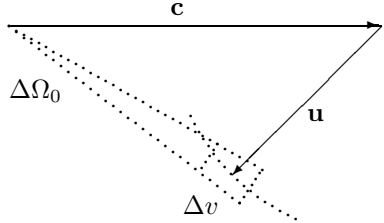
\begin{center} \begin{minipage}{10cm} { \vskip-0.3cm Figure~4: The relation 
between the velocity element $v^2\Delta v \Delta\Omega_0 $ and the velocity 
element $u^2 du d\Omega_c $.} \end{minipage} \end{center} \vspace{5pt} 

\noindent And, by investigating the situation in the velocity space shown in 
Fig.~4, the following relation can be found out: 
\begin{equation} \int_{\Delta v \Delta \Omega_0}u^2 du d\Omega_c \cdots 
\approx v^2\Delta v \Delta\Omega_0\cdots .
\end{equation}
Therefore, the first-order distribution function at the entry of the detector 
is 
\begin{equation} \label{final} f({\bf r}, {\bf v}, \Delta \Omega)=\frac{1}{v 
\Delta \Omega} \int_{-\Delta \Omega}d{\bf r}' \int_0^\infty dc 
\frac{8\sigma_c({\bf u}',{\bf u})g(r',c+u) g(r',c-u)}{u|{\bf r}-{\bf r}'|^2}, 
\end{equation}
in which, although $u =|{\bf c}-{\bf v}|$, $u<c$ needs to be ensured for the 
integral to make sense and the direction of $\bf u'$ is the same as that of 
${\bf c}'$ or ${\bf c}$. 

Since we have, at the beginning of this section, assumed $\Delta\Omega$ to be 
a finite solid-angle range, expression (\ref{final}) is nothing but the 
distribution function averaged over $\Delta\Omega$, which is, in a sense, 
still different from `the true and exact distribution function' there. In the 
next section, we shall come back to this issue and find out that the true and 
exact distribution function is actually beyond our reach. 

If interested, readers may test the resultant formalism with aid of realistic 
and computational experiments or develop it to cover more practical and more 
complicated cases. This is one of the major purposes of the present paper. 

\section{Discussion} 

In this section, we wish to further justify one of our introduced concepts, 
the distribution function averaged over finite solid-angle ranges of 
velocity. \vskip 10pt 

\hspace{30pt} \setlength{\unitlength}{0.018in} \vspace{0.5cm} 
\begin{picture}(100,100) \hspace{0.0cm} 

\put(30.797, 87.459){\circle*{1.0}}
\put(31.083, 86.931){\circle*{1.0}}
\put(31.369, 86.404){\circle*{1.0}}
\put(31.656, 85.877){\circle*{1.0}}
\put(31.942, 85.349){\circle*{1.0}}
\put(32.228, 84.822){\circle*{1.0}}
\put(32.514, 84.294){\circle*{1.0}}
\put(32.800, 83.767){\circle*{1.0}}
\put(33.044, 83.219){\circle*{1.0}}
\put(33.288, 82.671){\circle*{1.0}}
\put(33.532, 82.122){\circle*{1.0}}
\put(33.776, 81.574){\circle*{1.0}}
\put(34.020, 81.026){\circle*{1.0}}
\put(34.263, 80.478){\circle*{1.0}}
\put(34.507, 79.930){\circle*{1.0}}
\put(34.751, 79.381){\circle*{1.0}}
\put(34.951, 78.816){\circle*{1.0}}
\put(35.152, 78.250){\circle*{1.0}}
\put(35.352, 77.685){\circle*{1.0}}
\put(35.552, 77.119){\circle*{1.0}}
\put(35.753, 76.553){\circle*{1.0}}
\put(35.953, 75.988){\circle*{1.0}}
\put(36.153, 75.422){\circle*{1.0}}
\put(36.318, 74.845){\circle*{1.0}}
\put(36.483, 74.269){\circle*{1.0}}
\put(36.648, 73.692){\circle*{1.0}}
\put(36.813, 73.115){\circle*{1.0}}
\put(36.978, 72.538){\circle*{1.0}}
\put(37.143, 71.961){\circle*{1.0}}
\put(37.308, 71.384){\circle*{1.0}}
\put(37.472, 70.807){\circle*{1.0}}
\put(37.600, 70.221){\circle*{1.0}}
\put(37.728, 69.635){\circle*{1.0}}
\put(37.856, 69.049){\circle*{1.0}}
\put(37.984, 68.462){\circle*{1.0}}
\put(38.112, 67.876){\circle*{1.0}}
\put(38.240, 67.290){\circle*{1.0}}
\put(38.368, 66.704){\circle*{1.0}}
\put(38.496, 66.118){\circle*{1.0}}
\put(38.624, 65.531){\circle*{1.0}}
\put(38.713, 64.938){\circle*{1.0}}
\put(38.802, 64.345){\circle*{1.0}}
\put(38.890, 63.751){\circle*{1.0}}
\put(38.979, 63.158){\circle*{1.0}}
\put(39.068, 62.564){\circle*{1.0}}
\put(39.157, 61.971){\circle*{1.0}}
\put(39.245, 61.378){\circle*{1.0}}
\put(39.334, 60.784){\circle*{1.0}}
\put(39.389, 60.187){\circle*{1.0}}
\put(39.444, 59.589){\circle*{1.0}}
\put(39.500, 58.992){\circle*{1.0}}
\put(39.555, 58.394){\circle*{1.0}}
\put(39.610, 57.797){\circle*{1.0}}
\put(39.665, 57.199){\circle*{1.0}}
\put(39.720, 56.602){\circle*{1.0}}
\put(39.775, 56.005){\circle*{1.0}}
\put(39.830, 55.407){\circle*{1.0}}
\put(39.849, 54.807){\circle*{1.0}}
\put(39.868, 54.208){\circle*{1.0}}
\put(39.887, 53.608){\circle*{1.0}}
\put(39.906, 53.008){\circle*{1.0}}
\put(39.924, 52.409){\circle*{1.0}}
\put(39.943, 51.809){\circle*{1.0}}
\put(39.962, 51.209){\circle*{1.0}}
\put(39.981, 50.609){\circle*{1.0}}
\put(40.000, 50.010){\circle*{1.0}}
\put(39.982, 49.410){\circle*{1.0}}
\put(39.965, 48.810){\circle*{1.0}}
\put(39.948, 48.210){\circle*{1.0}}
\put(39.931, 47.611){\circle*{1.0}}
\put(39.913, 47.011){\circle*{1.0}}
\put(39.896, 46.411){\circle*{1.0}}
\put(39.879, 45.811){\circle*{1.0}}
\put(39.862, 45.212){\circle*{1.0}}
\put(39.812, 44.614){\circle*{1.0}}
\put(39.763, 44.016){\circle*{1.0}}
\put(39.713, 43.418){\circle*{1.0}}
\put(39.663, 42.820){\circle*{1.0}}
\put(39.614, 42.222){\circle*{1.0}}
\put(39.564, 41.624){\circle*{1.0}}
\put(39.515, 41.026){\circle*{1.0}}
\put(39.465, 40.428){\circle*{1.0}}
\put(39.415, 39.830){\circle*{1.0}}
\put(39.329, 39.236){\circle*{1.0}}
\put(39.243, 38.643){\circle*{1.0}}
\put(39.156, 38.049){\circle*{1.0}}
\put(39.070, 37.455){\circle*{1.0}}
\put(38.984, 36.861){\circle*{1.0}}
\put(38.898, 36.268){\circle*{1.0}}
\put(38.811, 35.674){\circle*{1.0}}
\put(38.725, 35.080){\circle*{1.0}}
\put(38.605, 34.492){\circle*{1.0}}
\put(38.484, 33.905){\circle*{1.0}}
\put(38.363, 33.317){\circle*{1.0}}
\put(38.243, 32.729){\circle*{1.0}}
\put(38.122, 32.141){\circle*{1.0}}
\put(38.002, 31.554){\circle*{1.0}}
\put(37.881, 30.966){\circle*{1.0}}
\put(37.761, 30.378){\circle*{1.0}}
\put(37.640, 29.790){\circle*{1.0}}
\put(37.479, 29.212){\circle*{1.0}}
\put(37.319, 28.634){\circle*{1.0}}
\put(37.158, 28.056){\circle*{1.0}}
\put(36.997, 27.478){\circle*{1.0}}
\put(36.836, 26.900){\circle*{1.0}}
\put(36.676, 26.322){\circle*{1.0}}
\put(36.515, 25.744){\circle*{1.0}}
\put(36.354, 25.166){\circle*{1.0}}
\put(36.155, 24.600){\circle*{1.0}}
\put(35.956, 24.034){\circle*{1.0}}
\put(35.756, 23.468){\circle*{1.0}}
\put(35.557, 22.902){\circle*{1.0}}
\put(35.358, 22.336){\circle*{1.0}}
\put(35.158, 21.770){\circle*{1.0}}
\put(34.959, 21.204){\circle*{1.0}}
\put(34.723, 20.653){\circle*{1.0}}
\put(34.487, 20.101){\circle*{1.0}}
\put(34.251, 19.549){\circle*{1.0}}
\put(34.015, 18.998){\circle*{1.0}}
\put(33.778, 18.446){\circle*{1.0}}
\put(33.542, 17.895){\circle*{1.0}}
\put(33.306, 17.343){\circle*{1.0}}
\put(33.070, 16.792){\circle*{1.0}}
\put(32.789, 16.261){\circle*{1.0}}
\put(32.508, 15.731){\circle*{1.0}}
\put(32.227, 15.201){\circle*{1.0}}
\put(31.946, 14.671){\circle*{1.0}}
\put(31.666, 14.141){\circle*{1.0}}
\put(65,50){\vector(1,0){50}}
\put(65,56.5){\vector(4,1){50}}
\put(65,43.5){\vector(4,-1){50}}
\multiput(80,90)(-2,2){3}{\vector(-1,-1){15}}
\put(101.5,54){\makebox(20,8)[l]{${\bf r}$}}
\put(16,42){\makebox(20,8)[r]{$\Delta_i S$}}
\put(100,56){\circle*{2}}
\put(38.5,50){\oval(3,5)}

\end{picture} 

\vspace{-28pt}\begin{center} \begin{minipage}{10cm} { \vskip-0.3cm Figure~5: 
Particles refleted by a surface.} \end{minipage} \end{center} 

Let's look at the path-invariance theorem expressed by (\ref{fm}) again. 
Although the path-invariance is not equivalent to the collisionless Boltzmann 
equation (see the remark at the end of Sect.~3), the theorem can still be 
thought of as a part of the conventional theory since it is formally the same 
as the collisionless Boltzmann equation and it is consistent with the usual 
notion that distribution functions of real gases are continuous or, at least, 
piecewise continuous. Then, a crucial question arises. Are all distribution 
functions of realistic gases continuous or, at least, piecewise continuous? 
If this is the case, we will be able, at least in principle, to determine the 
exact distribution function. To see what the real situation is, we examine 
the case illustrated in Fig.~5, where particles are moving toward a surface 
with a definite velocity (or, equivalently, let the surface move toward the 
particles). Since the surface is not uniform geometrically and physically, we 
need to, in order to formulate the reflected particles, divide it into many 
small surface elements, denoted by $\Delta_i S$ herein. It is easy to see 
that, corresponding to one of $\Delta_i S$ the reflected particles are like 
ones emitted from a point particle source and the distribution function, at a 
position ${\bf r}$ around the surface, is 
\begin{equation} \label{delta1} \Delta_i f = \frac{\eta_i({\bf v}) \Delta_i 
S }{4\pi |{\bf r}-{\bf r}_i|^2} \delta\left(\Omega_{\bf v}- \Omega_{{\bf 
r}{-\bf r}_i}\right), 
\end{equation}
where ${\bf r}_i$ represents the position vector of $\Delta_i S$ and 
$\eta_i({\bf v})$ is the reflection (emission) function that can be 
determined by experiments\cite{kogan}. As has been illustrated in Sect.~3, 
the distribution function expressed by (\ref{delta1}) is an infinitely thin 
`sting' at every spatial point, and the particle density along a particle 
path is no longer invariant. The total distribution function at ${\bf r}$ is 
\begin{equation} \label{delta2} f=\sum_i \Delta_i f = \sum_i 
\frac{\eta_i({\bf v}) \Delta_i S }{4\pi |{\bf r}-{\bf r}_i|^2} 
\delta\left(\Omega_{\bf v}- \Omega_{{\bf r}{-\bf r}_i}\right). 
\end{equation}
It is then found that, if the true and exact distribution function is indeed 
of our concern, the $\delta$-functions in (\ref{delta2}) will stay there and 
no regular functions can be used to replace them because each of 
$\Omega_{{\bf r}{-\bf r}_i}$ in it points in a distinctive direction, which 
means, in a more vivid language, the velocity distribution at every spatial 
point is shaped like an infinite number of infinitely thin stings (similar to 
functions dealt with in the studies of fractals). Without introducing the 
distribution function averaged over velocity solid-angle ranges, we will have 
great difficulty in setting up a calculable theory. \vspace{-33pt} 

\hspace{15pt} \setlength{\unitlength}{0.023in} \begin{picture}(200,105) 

\multiput(40,20)(40,0){2}{\line(0,1){15}}
\multiput(40,35)(22,0){2}{\line(1,0){18}}
\put(60,35){\line(1,5){6}}
\put(60,35){\line(-1,5){6}}
\multiput(60.5,33)(0.4,-2){8}{\circle*{0.7}}
\multiput(59.5,33)(-0.4,-2){8}{\circle*{0.7}}
\put(66,23){\makebox(35,8)[l]{$\Delta\Omega$}}
\put(60.5,20){\line(2,1){7}}

\multiput(60,73)(0.2,-2){6}{\circle*{0.7}}
\multiput(60,73)(-0.2,-2){6}{\circle*{0.7}}
\put(65,68){\makebox(35,8)[l]{$\Delta\Omega_0$}}
\put(60,64){\line(2,1){7}}
\put(61,35.5){\line(2,1){7}}
\put(69,38.5){\makebox(35,8)[l]{$\Delta S$}}
\put(133,35){\line(2,1){7}}
\put(141,38){\makebox(35,8)[l]{$\Delta S$}}

\multiput(110,20)(40,0){2}{\line(0,1){15}}
\multiput(110,35)(26,0){2}{\line(1,0){14}}

\multiput(124,35)(12,0){2}{\line(0,1){28}}
\multiput(126,35)(2,0){5}{\circle*{0.7}}

\multiput(130,35)(0.2,-2){9}{\circle*{0.7}}
\multiput(130,35)(-0.2,-2){9}{\circle*{0.7}}
\put(135,24){\makebox(35,8)[l]{$\Delta\Omega$}}
\put(130,21){\line(2,1){7}}

\put(130,70){\vector(0,-1){14}}
\put(126,49){\makebox(8,8)[c]{$\bf v$}}
\put(55,8){\makebox(10,8)[c]{$\bf (a)$}}
\put(125,8){\makebox(10,8)[c]{$\bf (b)$}}
\end{picture} 

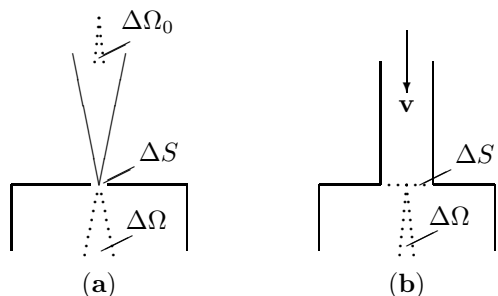
\vspace{-23pt}\begin{center} \begin{minipage}{10cm} {Figure~6: The effective 
region above the detector: (a) If $\Delta\Omega$ is fixed and $\Delta S$ 
shrinks to zero, it is cone-shaped. (b) If $\Delta S$ is fixed and 
$\Delta\Omega$ shrinks to zero, it is cylinder-shaped. } \end{minipage} 
\end{center} 

Treatment of particle-to-particle interactions is always an indispensable 
chapter of statistical mechanics. While referring readers to 
elsewhere\cite{chen1,chen2} for analyses of perplexing problems in the 
textbook methodology, we shall here comment on our own treatment of collision 
presented in the last section. At first glance, if $\Delta \Omega$ there were 
infinitely small, both the denominator and numerator of (\ref{final}) would 
tend to zero simultaneously and the total expression would remain finite and 
valid. A careful inspection, however, tells us that there are two `competing' 
quantities $\Delta\Omega$ and $\Delta S$. If we fix $\Delta \Omega$ and let 
$\Delta S$ approach zero, the effective region is a cone-shaped one, shown in 
Fig.~6a, and the related distribution function is, as has been formulated, 
the one averaged over the finite solid-angle range $\Delta\Omega$. Whereas, 
if we fix $\Delta S$ and let $\Delta \Omega$ approach zero, the situation is 
rather different: the effective region becomes a cylinder-shaped one, shown 
in Fig.~6b, and the related distribution function will be, if formulated, the 
one averaged over the finite spatial area $\Delta S$. In both the above 
situations, each of the results is related to an integration, each of the 
integrations is taken over a region infinitely-extended in space. (Even if 
more collisions are taken into account the above assertion still holds its 
significance since particles from the remote regions can still come freely in 
terms of probabilities.) There is no way to ensure the continuity of each 
formalism let alone the consistency between the two formalisms.

The discussions in this section clearly show that, in dealing with both 
collisionless dynamics and collisional dynamics, a finite range of certain 
variable (whether in the position space or in the velocity space) and the 
average over this range have to be employed. If the approach is inherently 
and completely that of differentiation, which means that the exact 
distribution function is of concern and every position and velocity range 
must shrink to zero, insurmountable conflicts surface automatically in one 
way or another. The repetitive appearances of conflict remind us of the 
situation in quantum mechanics where the accurate position and momentum 
cannot be determined simultaneously. Possible implication of this issue has 
yet to be explored. 

\section{Summary} We have proposed a gas as our working model and formulated 
a feasible method to calculate it. The formalism shows that real gases, at 
least certain types of them, can be calculated at the kinetic level. It is 
expected that such calculations will soon be compared to realistic or 
computational experiments. 

In this paper, collisional effects were investigated under the assumption 
that each particle involves, at most, one collision. If more collisions are 
taken into account, the trajectory of a particle will be very much like that 
of the Brownian motion or, in another sense, similar to that appearing in the 
logistic map of nonlinear studies. Although the extensions in this direction 
have been tried\cite{chen3,chen4}, it is not appropriate to comment on them 
before this work is admitted by more in this community. 

Helpful discussions with professors Hanying Guo, Ke Wu and Qiang Chen are 
greatly appreciated. This paper is supported by School of Science, BUAA, PRC. 

 \end{document}